\documentclass[12pt]{iopart}
\usepackage{epsfig}

\begin{document}

\title{STAR: Recent Results and Future Physics Program}

\author{Olga Barannikova {\it for the STAR Collaboration}}

\address{University of Illinois at Chicago, 
845 W. Taylor St., Chicago, IL 60607, USA}
\ead{barannik@uic.edu}
\begin{abstract}

Two major advantages of the STAR detector - uniform azimuthal acceptance complementing
extended pseudo-rapidity coverage, and the ability to identify a wide variety of the hadron
species in almost all kinematic ranges - have allowed us to address successfully a set of key
physics topics at RHIC. We  report here selected recent results from the STAR experiment,
including  insights on system size effects on medium properties, hadronization mechanisms, and partonic energy loss from triggered and non-triggered probes. In conclusion, we present  an outlook on the  STAR new physics program in upcoming years.

\end{abstract}


\section{Introduction}

The ongoing heavy ion physics program with the STAR detector at RHIC has  accumulated the wealth of experimental data,  stimulating important advances in the understanding of properties of hot strongly interacting matter  \cite{b5}.  Some of the focus areas currently under study are the thermodynamic and hydrodynamic properties of the  matter created,  in-medium effects on parton propagation, and hadronization mechanisms in QCD matter.
Uniform azimuthal acceptance and extended pseudo-rapidity coverage of the STAR detector is exceptionally suited for studies of jets and jet-medium interactions via triggered and non-triggered correlations. 
The data  provided  quantitative constrains on jet fragmentation and theoretical models of energy loss.  
Systematic studies of medium effects are advanced by analysis of system size effects on  tomographic probes.  Data collected with lighter Cu ions  have expanded those studies  by providing additional control over the system size variations. Complimentary data samples at lower beam energy of 62 GeV have also allowed  systematic studies as function of energy to be performed. 

\section{Particle Identification}
STAR full capabilities for particle identification in a wide kinematic range  have enabled new advances in more quantitative tests of QCD predictions. 
Different detectors provide complimentary capabilities and coverage. Pions, kaons, electrons and (anti)protons are identified via their specific ionization energy loss in the Time-projection Chamber (TPC) material at low and high transverse momenta ($p_T$). In the intermediate momentum range TPC results are enhanced by the Time-of-Flight (TOF) capabilities.  
Muons can be identified in the Muon Telescope Detector (MTD) prototype.
Photons at forward rapidities are measured  using the Photon Multiplicity Detector (PMD).
Weakly decaying strange and multi-strange hadrons are identified geometrically by their decay vertices with the TPC and TOF information on decay daughter particles. Short-lived resonances are reconstructed combinatorially by the invariant mass technique.

These capabilities have allowed measuring the yields and transverse momentum spectra addressing a variety of physics topics, including strangeness and heavy flavor production that are of topical interest for this conference. Strangeness enhancement was long thought as one of the signatures for the quark-gluon plasma formation \cite{strange}. Early data suggested that strangeness equilibration is achieved at RHIC and that initial collision geometry affects the strangeness systematics.  In STAR studies of the collision geometry effects on strangeness production is being accomplished by comprehensive analysis of various strange species ($K^{\pm}$,$K^0$, $\Lambda$, $\Xi$, $\phi$, $\Omega$) in light vs. heavy systems at all RHIC energies. The latest update on those developments is reported in \cite{Ant}.

New advances in heavy flavor sector afforded by STAR particle identification capabilities include the measurements of J/$\psi$ spectra and nuclear modification factors in Cu+Cu collisions, measurements of 
charm total cross-section for $pp$, d+Au, Au+Au and Cu+Cu systems, new constrains on bottom contribution to heavy flavor  from J/$\psi$-hadron and $e^{\pm}$-$D^0$ azimuthal correlation studies. These and other highlights on heavy quark measurements from STAR are reported in \cite{XZB}.    
The following two sections 
will focus on some of the new  results from triggered and non-triggered correlation analyses.

\section{Non-triggered Correlations}

Elliptic flow results have produced a lot of excitement from the beginning of RHIC program.  For the first time in heavy ion collisions the large $v_2$ values reported could be described by ideal hydrodynamics~\cite{A2}. 
Theoretical calculations successful in description of flow results indicated the values of the shear viscosity over the entropy ratio consistent with the expectations for strongly interacting medium~\cite{A4}.  Although recent work seem to question the applicability of ideal hydro~\cite{A}, most theories agree that the matter created at RHIC is extremely dense,  with the hydrodynamic description applicability limits extending to higher $p_T$ for more central collisions. Flow results provide strong indications for partonic nature of the medium:  constituent quark scaling observed for all measured hadron species (including 
$\Omega$ and $\phi$ with their small hadronic interaction cross-sections) points on relevance of partonic degrees of freedom at hadronization~\cite{flow}. 
 Similar results are observed in the Cu+Cu collisions although the overall $v_2$ values are lower at a given centrality for the Cu+Cu compared to Au+Au at the same energy.

New result reported recently by STAR  on non-triggered two-particle correlations provide independent means to extract  and cross-check azimuthal  anisotropy parameters, but also to characterize other  processes that induce relative angular correlations between the particles pairs. 
In this analysis di-hadron correlation formed with  no transverse momentum restrictions on either hadron  is used to study centrality evolution of characteristic structures in relative azimuth and pseudo-rapidity. It has been shown that the resulting correlation  can be decomposed into a few  structures, and  information  related to the individual sources can be extracted~\cite{minijets}.  
 Per-particle pair densities for various centrality bins of 200~GeV minimum bias Au+Au data sample are shown in Figure 1. 
\begin{figure}[!htb]
\centering
\vspace*{-0.5cm}
\includegraphics[width=0.95\textwidth]{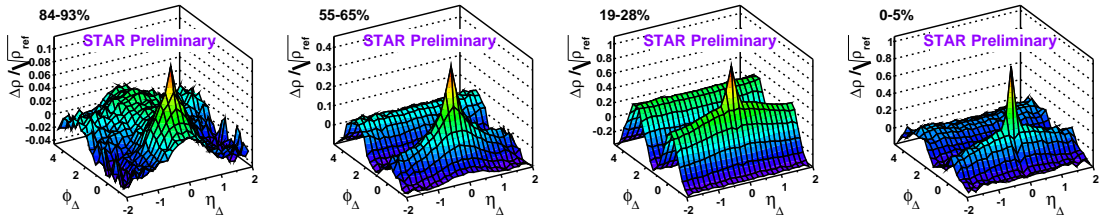}
\vspace*{-0.3cm}
\caption{ Evolution of non-triggered pair densities from  peripheral (left) to
central (right)  events in 200 GeV Au+Au collisions.} 
\end{figure}

Minijet contribution, measured as the same-side  two-dimensional Gaussian correlation structure, was found to exhibit a threshold behavior.  While approximate binary scaling  is observed from $pp$ and  most peripheral  Au+Au to mid-central collisions,  a sharp transition  is found for both 200 and 62 GeV Au+Au data  at about 55\% and 40\% event centrality, respectively. As shown in Figure 2, from this point the minijet peak amplitude exhibits rapid growth and the peak width in relative pseudo-rapidity broadens by more than a factor of 2.  Here the transverse density, calculated as  $\frac{dN_{Ch}/d\eta}{S}$, where $S$ is initial collision overlap area S from Glauber model,  is used as a centrality measure. Also shown in Figure 2 is the  total yield or "volume" of the minijet peak. It is estimated to include approximately one third of all produced particles in central collisions.  
\begin{figure}[!htb]
\centering
\vspace*{-0.5cm}
\includegraphics[width=0.75\textwidth]{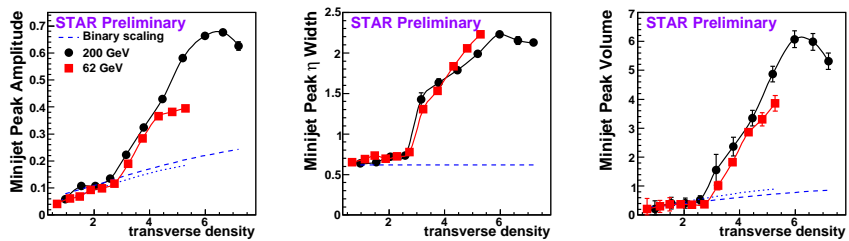}
\vspace*{-0.3cm}
\caption{ Evolution of the same-side  peak amplitude, relative pseudo-rapidity width, and volume extracted from non-triggered di-hadron correlations. Data 
shown as function transverse density, corresponding to the following centrality bins for each colliding energy: 84-93\%, 74-84\%, 65-74\%, 55-65\%, 46-55\%,
37-46\%, 28-37\%, 19-28\%, 9-19\%, 5-9\%, and 0-5\%.} 
\end{figure}

\section{Triggered Correlations}

Novel features have been also observed by STAR in the triggered correlations. Early results on the disappearance of away-side jet (in certain kinematic  range)  unambiguously established the presence of the jet quenching phenomena at RHIC \cite{b5}. 
Experimental observations are found to be consistent with theoretical expectations for partonic energy loss \cite{b6,b7}, prompting more studies of the mechanisms of jet-medium interactions. 

Two-dimensional study of the correlation structures for charged hadrons associated with a high-$p_T$ trigger shows shape modifications in both $\Delta\phi$ and $\Delta\eta$ dimensions compared to a $pp$ reference.
Within a two-component approach the intermediate $p_T$ structure can be decomposed into a  jet-like part, and the so-called "ridge". The jet part appears similar in shape to the structures seen in $pp$ data, while "ridge" resembles  properties of the bulk medium. Left panel of Figure 3 illustrates the ridge - long range $\Delta\eta$ structure   accompanying the jet  peak in central 200 GeV Au+Au collisions\cite{ridge}.  
Jet-medium interaction is likely to be the  origin of the ridge formation and modification of the jet-like peaks.
Di-hadron correlations with identified hadrons  provide additional means to study parton propagation in  the medium. Measurements of identified particle  spectra and ratios for associated hadrons in the jet cones  and ridges could address the issues of color-charge dependence of energy loss, presence of meson/baryon effects and interplay between various hadronization scenarios. 
Strong enhancement of baryon production at intermediate momentum range in inclusive measurements was theorized as an evidence for quark coalescence. However, in azimuthal correlations with the hyperon-triggers  significant jet-like structures were observed for all species of hyperons (including $\Omega$)~\cite{jana}, which  contradicts initial expectations from  coalescence picture.
 Associated hadron yields in  (multi)strange hadron correlations show no dependence  on strangeness content. Following those observations, a medium response to the propagating parton or "phantom jet" was proposed to accommodate the data within the coalescence picture~\cite{hwa}. This theoretical explanation preserves the correlation in azimuthal space and produces ridge-like structure in pseudo-rapidity. To confront the models and further the understanding of physics mechanisms producing the ridge, it was essential to extend the correlation studies  into $\Delta\eta$-$\Delta\phi$ space separately  for baryons and mesons. 

Preliminary STAR results show significant differences in strange (central panel of Figure 3) and non-strange (Figure 3, right) baryon to meson ratios measured in jet part around a 4 GeV/$c$ triggers, with respect to the ridge part. The jet-part ratios for $p/\pi$ and $\Lambda/K^0$ are found to be significantly lower in central Au+Au collisions compared to non-triggered distributions, and are consistent with the inclusive ratios from 200 GeV $pp$ data~\cite{bedangaQM}. At the same time, the ridge ratios  suggest relative baryon enhancement for  both strange and non-strange species, which might be related to direct fragmentation of the gluons,  radiated of the hard-scattered quark, or their subsequent recombination with the medium partons.
Higher statistics A+A data samples, expected to be collected in future runs, are required to address ridge hadro-chemistry in more details.
\begin{figure}[!htb]
\vspace*{-0.2cm}
\begin{center}
\includegraphics*[height=3.2cm]{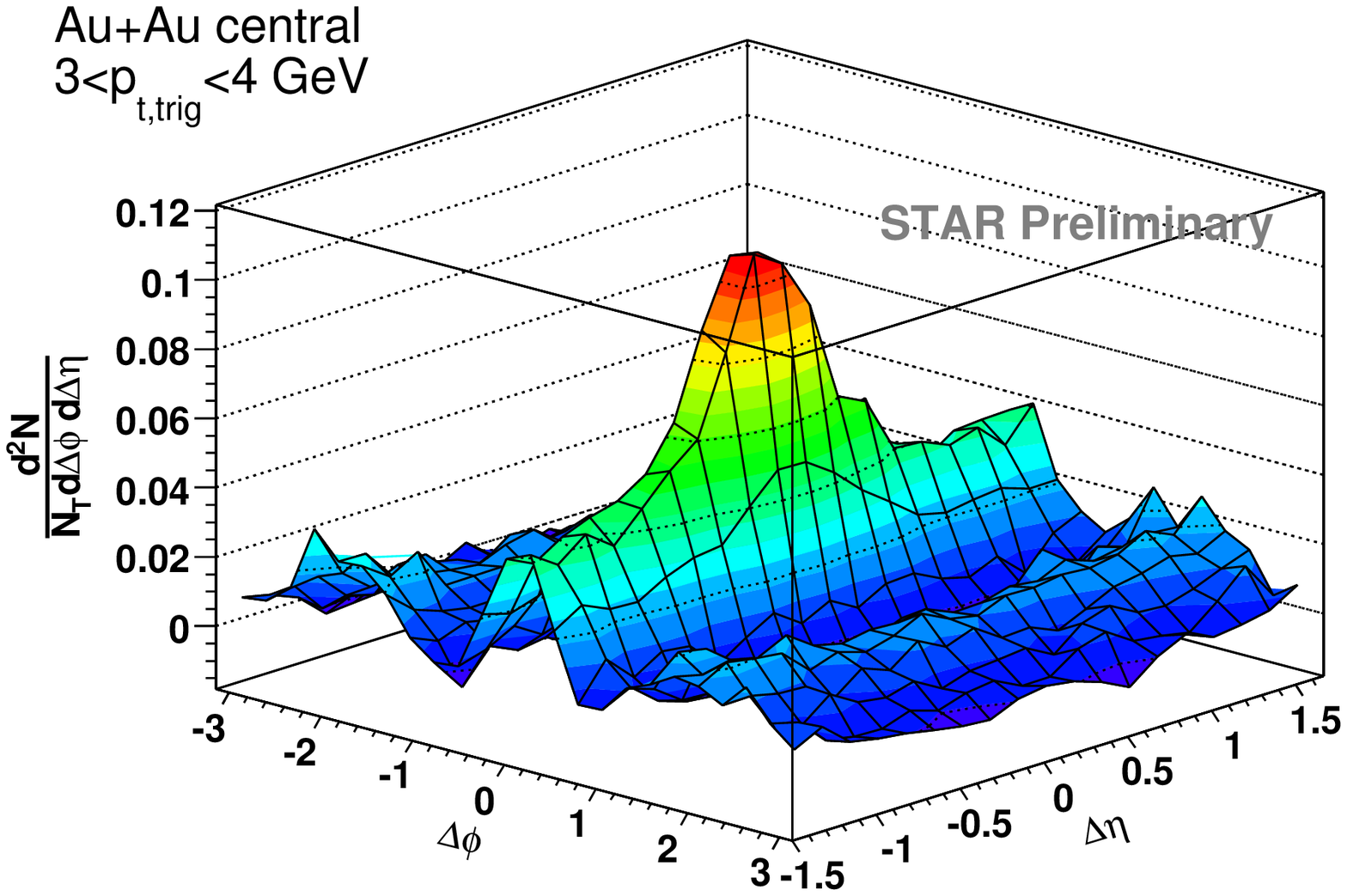} 
\includegraphics*[height=3.2cm]{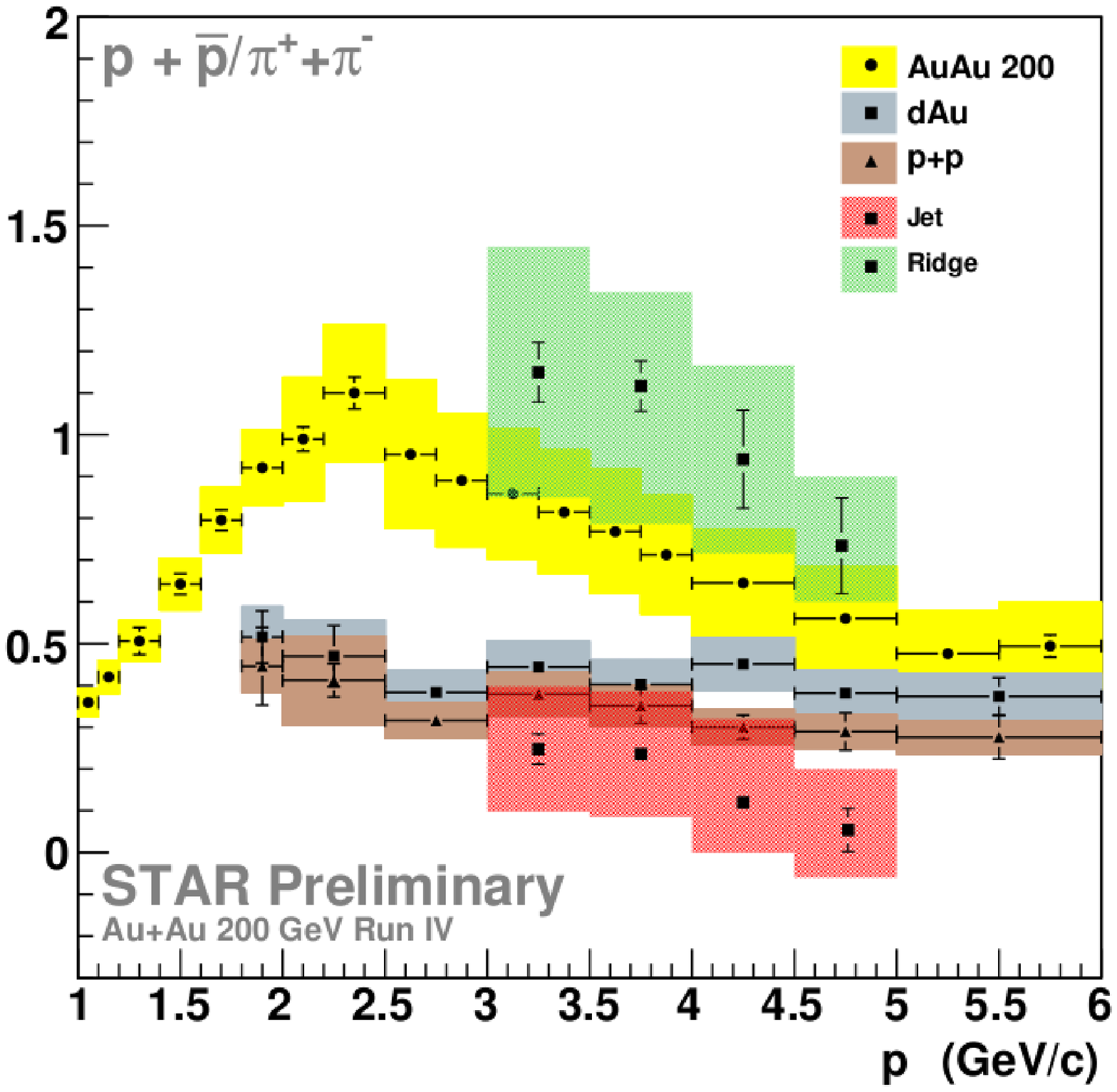} 
\includegraphics*[height=3.2cm]{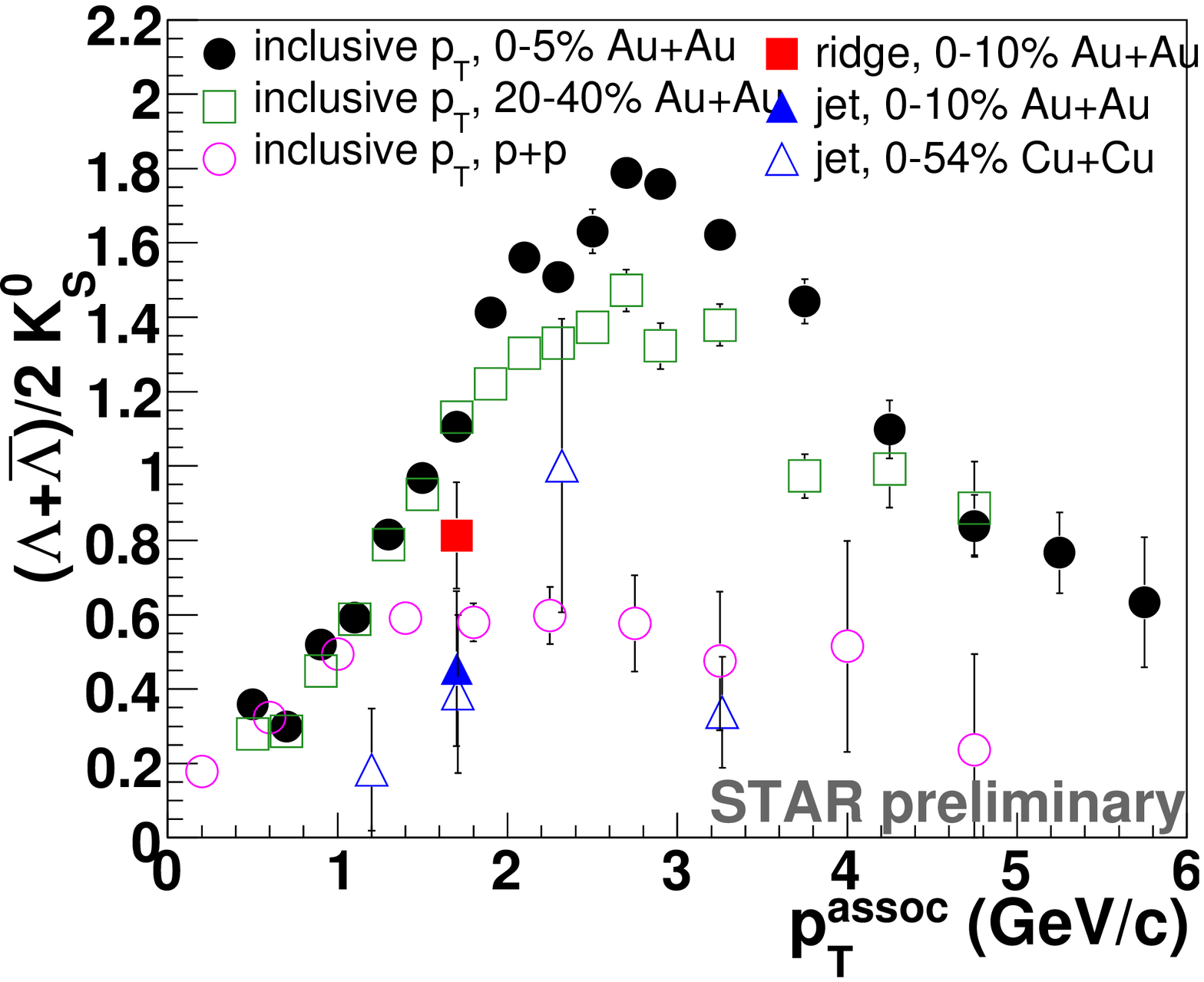} 
\end{center}
\vspace*{-0.3cm}
\caption{Left panel:  $\Delta\eta$-$\Delta\phi$ di-hadron correlation  in 200 GeV 12\% central Au+Au collisions for hadronic triggers with $p_T^{trig}>4$~GeV/$c$ and associate hadrons with $2<p_T^{trig}<4$~GeV/$c$. 
$\Lambda/K^0$ (center) and $p/\pi$ (right) ratios as function of $p_T$  for inclusive data, and ridge and jet parts in  200 GeV A+A data. Shown for comparison are the inclusive ratios from 200 GeV $pp$ and/or d+Au collisions. }
\end{figure}

To add yet another dimension to the ridge studies, one can bring in additional particle to the correlation to explore internal structure of the observed peak.  Preliminary results of three-particle triggered correlation analysis for unidentified hadrons are presented in Figure 4 for minimum bias d+Au, peripheral  and central Au+Au collisions  at 200 GeV.  A  jet-like peak is observed in all three correlations at small relative pseudo-rapidity values. In addition,  a uniform distribution of  associated yields is observed in Au+Au collisions, particularly, for most central events. This observation is quite puzzling for ridge phenomenology, contradicting  to  model predictions~\cite{r1,r2,r3}. 
Specifically we note the  absence of axial  strips in the correlation structure. Axial strips in three-particle correlation should be formed  by associate hadron pairs, in which one of the hadrons is coming  from the cone-part, and the other one - from the ridge. Absence of these strips thus 
 suggests complete decoupling of jet and ridge events.   Albeit  the current statistic limitations, these early results have already provided more stringent constraints on the interpretation of ridge formation.
\begin{figure}[!htb]
\centering
\vspace*{-0.5cm}
\includegraphics[width=0.95\textwidth]{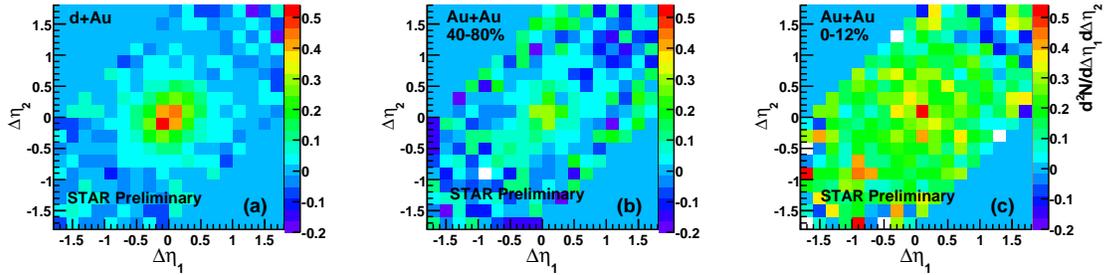}
\vspace*{-0.3cm}
\caption{ Pseudo-rapidity 3-particle correlations for associate hadrons with the small azimuthal distance to the trigger. Minimum bias d+Au (left), peripheral Au+Au (center) and central Au+Au (right) 
200 GeV collisions are shown. } 
\end{figure}

Full jet reconstruction could provide new means to measure the energy loss of propagating partons. Experimental techniques for such reconstruction have been successfully used  in elementary collision experiments. Recently full jet reconstruction was attempted for the first time for heavy ion events~\cite{Joern}. A combination of  charged hadron measurements from the TPC  and neutral energy measurements from Electro-magnetic calorimeter were utilized in STAR for this analysis. 
Left panel of Figure 5 depicts a 47~GeV single jet from 20\% central high-tower triggered (HT) Au+Au data of 200 GeV.  The aim of this analysis is to quantify the jet energy loss by detailed studies of possible modification in the fragmentation functions due to quenching effects in Au+Au with respect to $pp$ reference measurements.
Preliminary results from this study are shown in the right panel of Figure 5, where fragmentation functions, defined as $\xi=ln(p_{T,jet}^{rec}/p_T^{hadron})$, are compared for two colliding systems. 
No modification in the fragmentation functions in excess of  10 - 20\% is observed within the uncertainties from utilizing the current jet finding algorithms and background subtraction procedures . This preliminary conclusion depends heavily on the assumptions of Pythia fragmentation, and could be additionally affected by the bias induced in the jet-sample by applying background-suppressing analysis cuts. 
This work will be continued with the anticipated higher statistics Au+Au data in the coming years.

\begin{figure}[!htb]
\vspace*{-0.2cm}
\begin{center}
\includegraphics*[height=4.0cm]{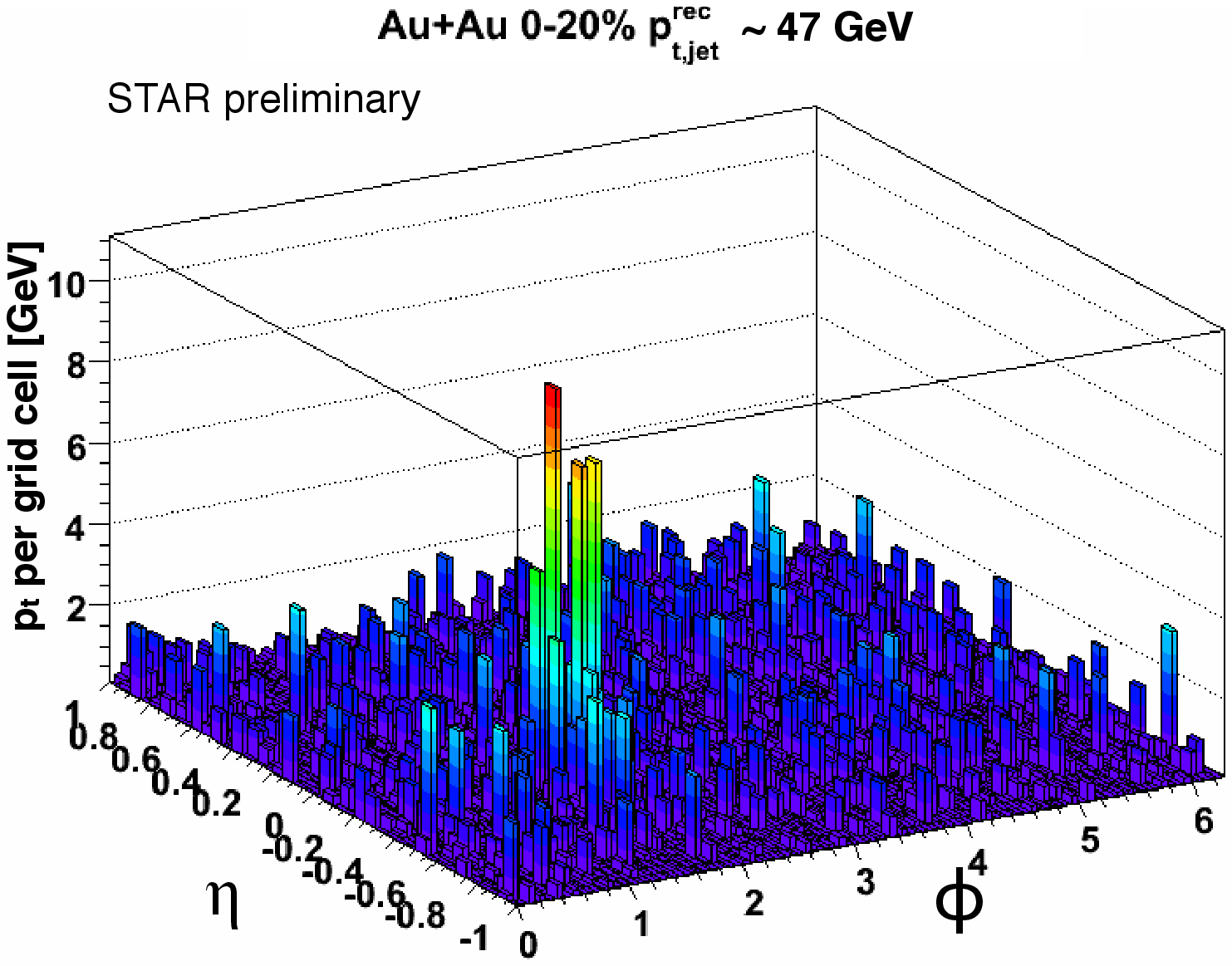} 
\includegraphics*[height=3.6cm]{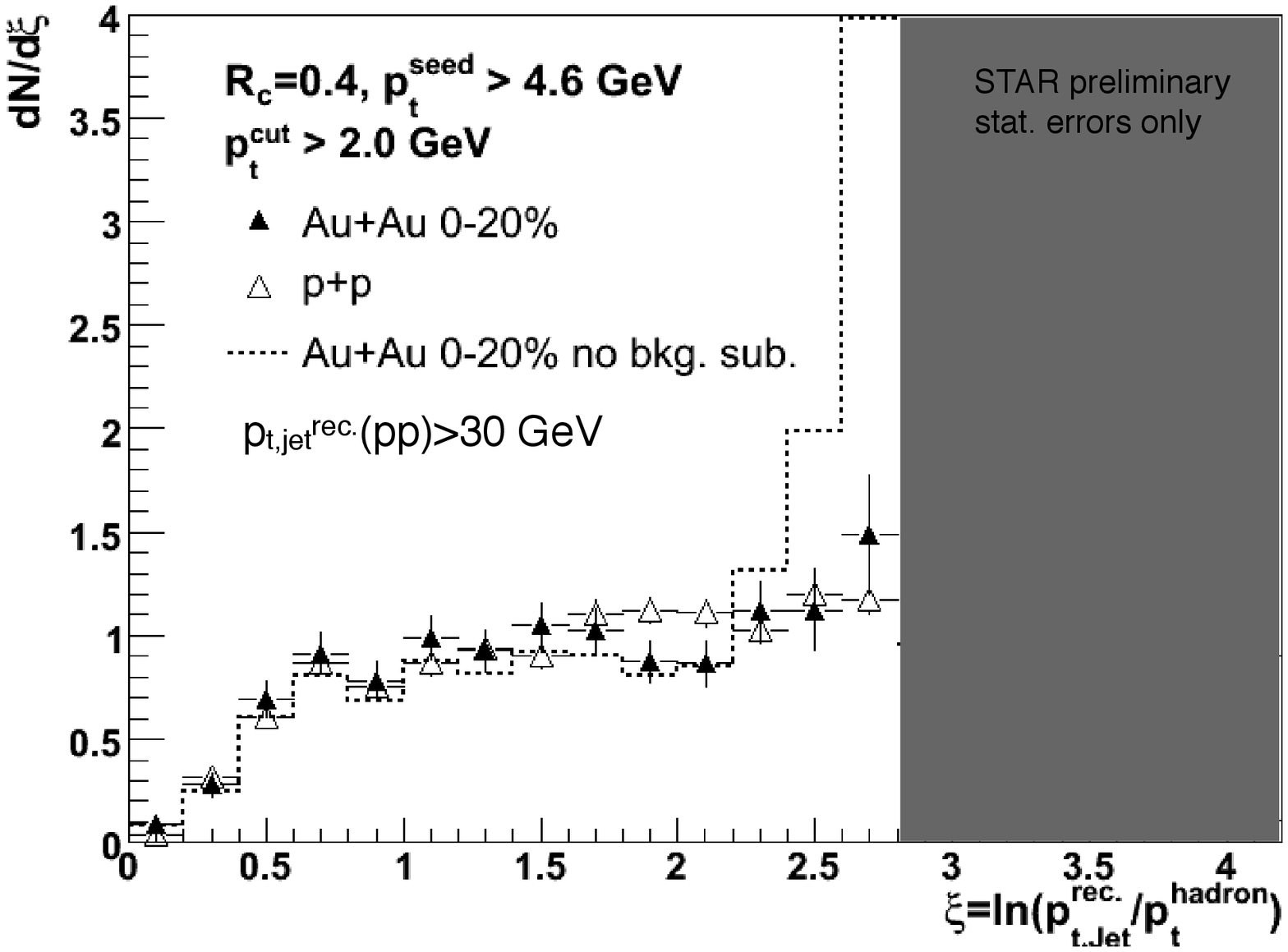} 
\end{center}
\vspace*{-0.3cm}
\caption{
Left panel: 47~GeV single jet with from 200 GeV 20\% central HT  Au+Au collisions. Right panel: Fragmentation function in 200 GeV 20\% central HT Au+Au collisions compared with the corresponding $pp$ data. Shaded area indicates the region where  Au+Au measurement are  dominated by the underlying background.}
\end{figure}

\section{Beam Energy Scan}

New physics focus will soon be added to the existing experimental program at RHIC~\cite{GO}. 
The compelling evidences gathered  for the new form of matter created in heavy ion collisions, called for more explorations of the QCD phase diagram.  The phase diagram, defined  in terms of  temperature (T) with respect to the baryon chemical potential ($\mu_B$), is believed to contain a number of phase boundaries, including the one that separates QGP partonic degrees of freedom  from the ordinary (hadronic gas) matter.
Lattice QCD (LQCD) predicts that in zero-$\mu_B$ limit the phase transition occurs as a smooth cross-over, while at higher $\mu_B$ values it is expected to be a first order transition. A Critical Point (CP) is expected  on a phase diagram, marking the location where QGP phase transition will change from a cross-over to a first order transition. In the absence of analytical solution various approximation techniques are used in the theoretical calculations, with variations of  lattice spacing, assumptions of quark masses and flavors, etc~\cite{Misha}. The resulting predictions with regard to the exact CP location   vary considerably.
 In the coming years STAR will engage in the Beam Energy Scan program, continuing years of research at AGS, SPS, and  RHIC itself.
To locate the Critical Point the Energy Scan program will vary incident energies to achieve desirable variations of  T and $\mu_B$.  Decreasing the center-of-mass energies leads to formation of colder matter with higher baryon potential, allowing one to move across the phase map.  

With the detailed scan  STAR will search for various CP signatures, as well as 
 investigate the evolution of medium properties indicative of the non-hadronic matter. Novel phenomena, such as constituent quark scaling of the elliptic flow,   hadron suppression at high-$p_T$ sigh signal the onset of deconfinment. From theory, the divergences are expected for the baryon number and the net charge susceptibilities, as well as for the specific heat near the Critical Point. Such divergences might lead to large  fluctuations  in a number of physical parameters, including  baryon number density, quark number, and charge, which could be observed experimentally in hadronic yields, net-charge and mean-$p_T$ fluctuations, as well as particle ratios ($K/\pi$, $p/\pi$, etc).  Previous experimental results might already provide some hints regarding the CP location. One such example is the excitation function for the $K^+/\pi$ ratio, that shows a drastic change-over near the center-of-mass energy of 8 GeV (Figure 6).
It is suggested that the "horn" feature could be linked directly to the QCD phase transition~\cite{Bnew}.
 The RHIC Energy Scan program will span a range of energies going slightly below that, covering a suggested region of  (T,$\mu_B$) values from most theoretical expectations.
STAR experiment is uniquely suited for the search armed not only by aforementioned PID capabilities, further enhanced by the completion of the full TOF barrel, but also by full azimuthal coverage with large acceptance, which is   essential for all fluctuation measurements.  The physics impact of fluctuation measurements,  for example in mean-$p_T$,  has  been already shown for the available RHIC energies,   providing essential baseline for the low energy scan.

\begin{figure}[!htb]
\centering
\vspace*{-0.5cm}
\includegraphics[width=0.4\textwidth]{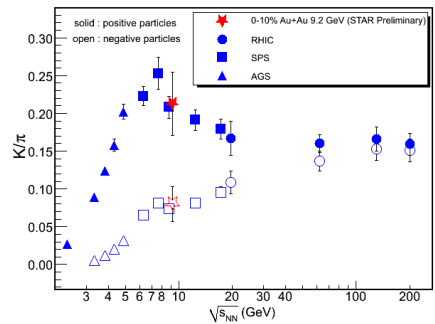}
\vspace*{-0.3cm}
\caption{ Charged kaon to  pion ratio as a function of center-of-mass
energy. Solid symbols show the $K^+/\pi^+$ ratio; the $K^-/\pi^-$ is shown with the open ones. Star symbols depict first results for mid-rapidity kaon to pion ratios from  9.2 GeV Au+Au collisions.} 
\end{figure}

The energy scan has essentially begun in STAR with the systematic studies of all available RHIC energies: 200, 130, 62.4 and 19.6 GeV.  In the most recent run (Run 8) an engineering study was undertaken by the Collider and Accelerator Department in conjunction with the Energy Scan program.  During this study, the accelerator physicists  demonstrated successful developments on injection and collisions of Au ions at  $\sqrt{s_{NN}}=9.2$ ~GeV.  STAR was able to record the first low energy collisions, collecting about 3000 good events suitable for physics analysis. Despite such limited statistics,  first results for identified hadron spectra, azimuthal anisotropy measurements and pion interferometry were already obtained~\cite{Lokesh}.  New data points added to the studies of the $K/\pi$ ratio systematics illustrate these achievements  in Figure 6 . With STAR's proven performance at low energy we are looking forward for new data in upcoming runs to further explore the QCD phase diagram and pin-point the location of the Critical Point.

\section{Conclusions and Outlook}

We have discussed selected recent results from ongoing correlation analyses from STAR experiment.  
The results on ridge properties from triggered and non-triggered correlation studies were highlighted.
The large baryon to meson ratios,  the uniformity of the ridge yield  in  $\Delta\eta$-$\Delta\eta$ space on event-by-event basis is suggestive of the bulk matter being the ridge origin.  
First attempts towards full jet reconstruction in heavy ion collisions hold a lot of potential  to provide further constraints on energy loss in QCD medium. The data sample currently available is limited in statistics to fully exploit this technique. 
In the coming years, STAR will continue collecting top energy data for $pp$ and Au+Au collisions,  providing the possibilities for more differential analyses. 

Another direction of STAR physics program is to partake in the RHIC Energy Scan. We will explore the QCD phase diagram in search for the onset of deconfinement and the location of Critical Point. STAR has demonstrated the readiness to deliver physics results for the program with the analysis of the recent low energy test run.

\end{document}